\documentclass[aps,twocolumn,showpacs,superscriptaddress]{revtex4-1}  
\usepackage{graphicx}  
\usepackage{dcolumn}   
\usepackage{bm}        
\usepackage{amssymb}   
\usepackage{color}

\newcommand{\xx}{\mathbf{x}}

\newcommand{\eqref}[1]{(\ref{#1})}

\begin{document}


\title{Impact of environmental colored noise in single-species population dynamics} 
\author{Tommaso~Spanio} \affiliation{Instituto Carlos I de F\'isica Te\'orica y Computacional and Departamento Electromagnetismo y F\'isica de la Materia, Universidad de Granada, 18071 Granada, Spain} \affiliation{Dipartimento di Fisica `G. Galilei' and CNISM, INFN, Universit\`a di Padova, Via Marzolo 8, 35131 Padova, Italy} \author{Jorge~Hidalgo} \email{hidalgo@pd.infn.it}
\affiliation{Dipartimento di Fisica `G. Galilei' and CNISM, INFN, Universit\`a di Padova, Via Marzolo 8, 35131 Padova, Italy}
\author{Miguel A.~Mu\~noz}
\affiliation{Instituto Carlos I de F\'isica Te\'orica y Computacional and Departamento Electromagnetismo y F\'isica de la Materia, Universidad de Granada, 18071 Granada, Spain}

\date{\today}

\begin{abstract}
Variability on the external conditions has important consequences for the dynamics and organization of biological systems, and, in many cases, the characteristic timescale of environmental changes as well as their correlations play a fundamental role in the way living systems adapt and respond to it. A proper mathematical approach to understand population dynamic, thus, requires of approaches more refined than e.g. simple white-noise approximations. To shed further light onto this problem, in this paper we propose a unifying framework based on different analytical and numerical tools available to deal with ``colored'' environmental noise.  In particular, we employ a ``unified colored noise approximation'' to map the original problem into an effective one with white noise, and then we apply a standard path integral approach to gain analytical understanding.  For the sake of specificity, we present our approach using as a guideline a variation of the contact process --which can also be seen as a
birth-death process of the Malthus-Verhulst class-- where the propagation/birth rate varies stochastically in time. Our approach allows us to tackle in a systematic manner some of the relevant questions concerning population dynamics under environmental variability such as, for instance, determining the stationary population density, establishing the conditions under which a population may become extinct, and estimating extinction times. More in general, we put the focus on the emerging phase diagram and its possible phase transitions, underlying how these are affected by the presence of environmental noise time-correlations.
 \end{abstract}

\maketitle

\section{Introduction}
\label{sec:intro}
Population dynamics is a core matter in the modeling of ecological communities, genetics, and epidemics \cite{May-eco,Kimura,May-epi}. Combined with the increasing volume of experimental available (big) data, it constitutes a fundamental tool to shed light into the laws governing complex communities of living systems \cite{Bialek2012}.  The traditional approach to population dynamics consists in the analysis of coupled deterministic equations describing the evolution of species abundances in a given community \cite{Murray}. This procedure --whose outcome is not necessarily simple \cite{May-paradox,Cordero2014}-- is adequate in many cases.  However, deterministic approaches neglect the effect of fluctuations, and these are now acknowledged to be both inherent and essential to the organization of communities of living systems \cite{RIDOLFI}. On the one hand, the discreteness and finiteness of populations lead to \emph{demographic noise}; which has been shown to be responsible of a wealth of non-trivial phenomena such as the emergence of complex statistical patterns in neutral communities \cite{DODORICO,Azaele2016}, quasi-periodic oscillations in prey-predator systems \cite{McKane2005}, species formation \cite{Rogers2012}, and others \cite{Constable2016,Leibler2017,Hidalgo2015}.  On the other hand, populations are strongly affected by fluctuations in external conditions \cite{Dunson1991, Pearman2008}, which in most of the cases are highly unpredictable. This source of stochasticity, usually called \emph{environmental noise}, can have important consequences for e.g. ecosystem stability \cite{Lewontin1969, Chevin2010} and evolutionary dynamics \cite{Levins1968, Frank1990, Ashcroft2014, Vergassola2015,Leibler2005}, and fosters species coexistence \cite{Chesson2000,Kalyuzhny2015, Nadav2016-1, Hidalgo2017}.

Remarkably, theoretical and empirical evidence reveals that these phenomena strongly rely on a specific interplay between the characteristic timescale of environmental variations and the intrinsic timescale of the dynamics \cite{Vasseur,Gonzalez,Nadav2016-1,Massie,Hidalgo2017}. Owing to this, theoretical approaches have to be constructed beyond simple white-noise approximations, i.e. including ``colored'' (time-correlated) noise \cite{halley1996, inchausti2002}.  The question of how environmental colored noise affects population extinction has been widely studied in the literature, and there are contrasting positions on whether environmental fluctuations increase or decrease the risk of extinction, as this may actually depend on subtle differences of the underlying dynamics as well as the actual ``color'' of the fluctuations \cite{lawton1988, ripa1996, petchey1997, cuddington1999, johst1997, schwager2006, Kamenev}. Of particular interest for our analyses here is the remarkable work of Kamenev {\it et al.} \cite{Kamenev}, who analyzed a logistic growth population-dynamic model in which birth and death rates fluctuate in time, showing that, depending on the interplay between the system size and the temporal scale of the environment, the model exhibits qualitatively different functional dependencies of the mean extinction time with the system size.

In this paper, we bring the question of how time-correlated environmental noise affects population dynamics to the context of phase transitions, and analyze in detail one of the most standard models in the study of population dynamics, the Contact Process \cite{Marro,Liggett} --which can also be described as a birth-death process of the Malthus-Verhulst class-- in the presence of time-correlated environmental variability \cite{TGP,TGP2,vojta2015,vojta2016,Jensen}.  To study this model we employ the so-called ``unified colored-noise approximation'' which is exact in the limits of very large and very short correlation times \cite{UCNA, Assaf2010}, and study the resulting effective (white-noise) problem employing a standard path integral approach; analytical results are tested against direct computational simulations obtained employing a very careful numerical analysis scheme \cite{Anderson}.  Using this combined approach, we scrutinize the model phase diagram and identify the parameters for which the population becomes extinct with certainty and those for which the population survives, as well as the threshold separating them, and how the resulting phase transition depends on environmental-noise time correlations. As we will show, the phase diagram becomes much richer in this case than in its noiseless counterpart.

From a broader perspective, our study provides a simple and general framework for the analysis of population dynamics with colored noise, blending together several analytical \cite{UCNA, Assaf2010} and numerical \cite{Anderson} tools already available in the literature, which can be straightforwardly implemented to other similar scenarios beyond the case presented here.

\section{The Contact Process}
\begin{figure}[bt] \includegraphics[width=\columnwidth]{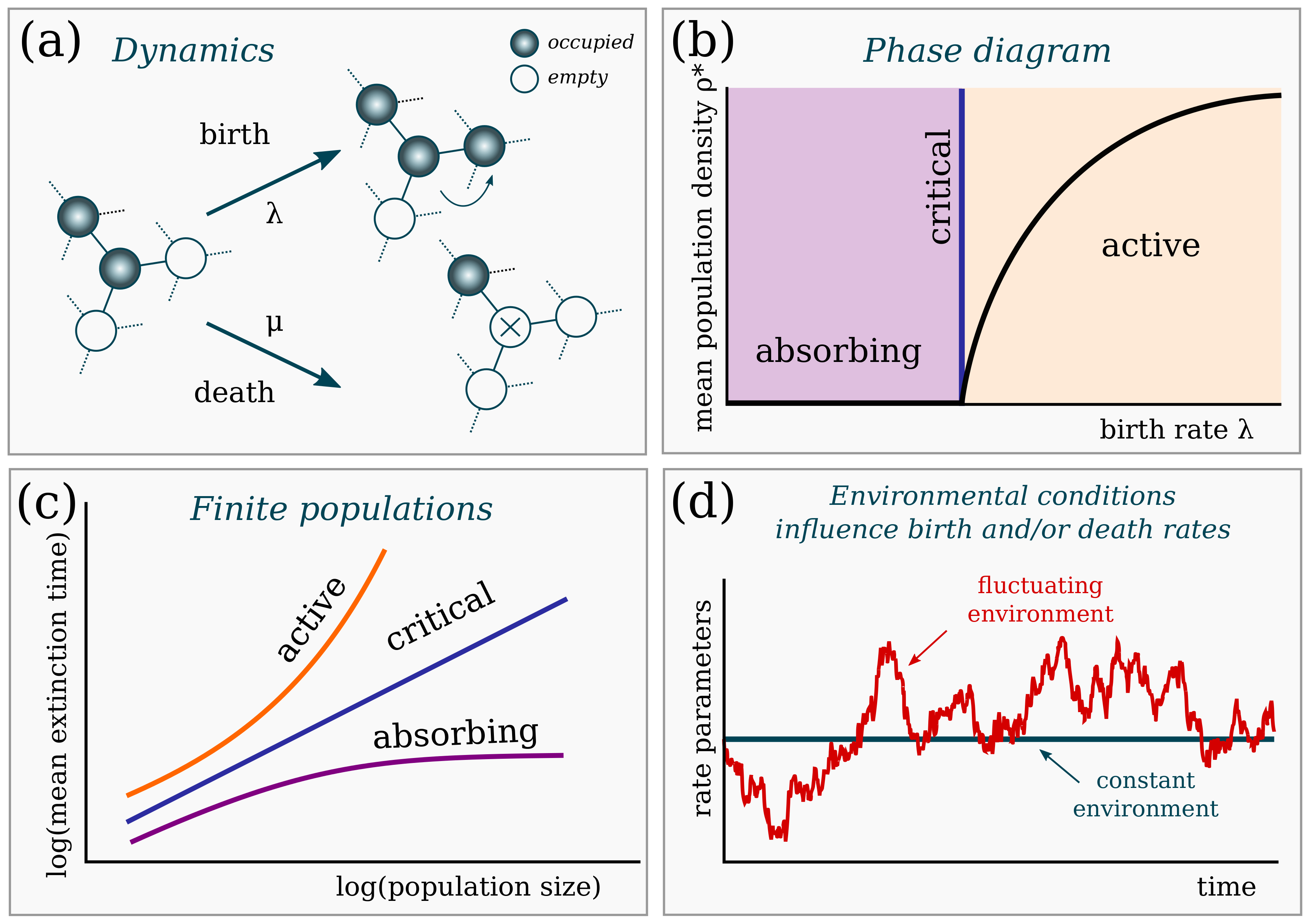} \caption{
\textbf{Single-species dynamics with extinction.} \textbf{(a)} A community of individuals grow under limited conditions with the dynamics of the Contact Process running on any given network.  Each of the $N$ nodes/locations in the network can be either occupied by up to one individual (active node) or remain empty (inactive node). Individuals can reproduce at empty neighboring nodes a at rate $\lambda$ and, also die and be removed from the community at rate $\mu$. \textbf{(b)} The model exhibits different behaviors depending on parameter values: for low values of the reproduction rate, birth processes do not compensate deaths, and any population becomes extinct on the long term (absorbing phase). After a threshold value, population density exhibits a non-zero stationary value (active phase). Both regimes are separated by a critical point of a continuous phase transition. \textbf{(c)} However, any finite population eventually becomes extinct due to demographic fluctuations. Phases can be then distinguished looking at the scaling of the mean-extinction time with the system size, which is logarithmic in the absorbing phase and exponential in the active phase, while it becomes a power-law just at the critical point. \textbf{(d)} In the simplest scenario, the rate at which individuals reproduce (and similarly for the death rate) can be considered as a constant parameter (blue line). However, variability on the external conditions such as the temperature, humidity, pH, ... may strongly influence rate parameters (red line). What is the impact of environmental variability in the previous panels?}
\label{fig:cp}
\end{figure}

The Contact Process (CP) \cite{Marro,Liggett} is a prototypical model for the study of population dynamics with extinction, with applications in different fields such as in epidemic spreading \cite{Vespignani2015}, ecology \cite{tilman1997}, and propagation of neural activity \cite{Moretti}. We use this simple model as a guideline here, but results are easily generalizable to other similar models for population or spreading dynamics.
In the CP (see Fig. \ref{fig:cp}), nodes in a given network (e.g. a square lattice) can be either occupied/active or vacant/inactive. Active nodes produce new offspring at neighboring empty sites at rate $\lambda$, and
can also die and be removed from the community at rate $\mu$. The total system size $N$ is fixed, representing limitation of space/resources, imposing an upper bound on the active population size. For the sake of simplicity, we neglect spatial effects and restrict our analysis to the simplest case of a well-mixed community (or, equivalently, a fully connected network).  At each time $t$, the state of the system is determined by the total number of active sites, $n(t)$, or equivalently, the population density, $\rho(t)=n(t)/N$. For very large populations, demographic fluctuations can be neglected, and the dynamics of $\rho$ becomes deterministic \cite{Marro, Liggett}: 
\begin{equation}
 \dot{\rho}(t) = \lambda \rho(t) (1-\rho(t)) -\mu \rho(t).
 \label{eq:deterministic}
\end{equation}

The stationary density, $\rho^*$ (see Fig. \ref{fig:cp}b) is either $\rho^*=0$ (the so-called ``absorbing'' state) if
$\lambda<\lambda_c=\mu$, i.e. if births do not balance deaths and the population progressively shrinks, leading to extinction, or $\rho^*=1-\mu/\lambda$ (``active'' phase) if $\lambda>\lambda_c$ and the population survives indefinitely. A ``critical'' point, $\lambda=\lambda_c$, separates the absorbing from the active phase; this value represents the extinction threshold and is a fundamental parameter of our forthcoming analysis.

In finite systems, demographic fluctuations can drive population extinction even when parameters correspond to the active phase \cite{Gardiner}; as a matter of fact, the only ``truly'' stable solution in the long term is the absorbing state in such a case. Still, it is possible to characterize the phases of a finite population by the mean time to reach extinction, $T$, as a function of the system size, $N$. Different functional dependencies emerge for each of the phases of the CP \cite{Henkel,TGP} (see Fig. \ref{fig:cp}): $T$ scales logarithmically with the system size in the absorbing phase ($\lambda<\lambda_c$), while it increases exponentially in the active phase ($\lambda>\lambda_c$) --meaning that extinctions become extremely rare for sufficiently large populations in the active phase-- and scales as a power-law right at the critical point, $\lambda=\lambda_c$. Thus, in a nutshell, the CP represents a prototypical paradigm to analyze single-species communities with extinction. Such a dynamics can be characterized by means of \textit{i)} the phase diagram, which describes the stationary state of the system as a function of the parameters, \textit{ii)} the critical point, representing the extinction threshold, and \textit{iii)} the scaling of the mean-extinction time with the system size, as a proxy for population stability.  In the next section we analyze how time-correlated environmental noise changes each of these elements and how this depends on the environmental auto-correlation time.

\subsection{Contact Process with environmental noise}

For the sake of simplicity, we assume that the environment influences homogeneously the population (i.e. demographic rates are global variables) and it does so by affecting only one parameter, that here we take to be the birth rate, leaving all other parameters unchanged (other choices are possible but they do not significantly affect the forthcoming results. The state of the environment is encoded in a time-dependent variable, $\epsilon(t)$, assumed to be independent of the state of system, so that $\lambda \rightarrow \lambda(t)=\bar\lambda + \sigma \epsilon(t)$, where $\bar\lambda$ is the mean value, $\sigma$ is a constant, and $\epsilon(t)$ follows an Ornstein-Uhlenbeck process \cite{Gardiner} (see Fig. \ref{fig:cp}): 
\begin{equation}
 \dot \epsilon = -\frac{1}{\tau}\epsilon + \sqrt{\frac{2}{\tau}} \xi(t).
 \label{eq:OU}
\end{equation}
where $\xi(t)$ is a zero-mean Gaussian white noise with $\langle \xi(t) \xi(t') \rangle = \delta(t-t')$.  From Eq. \eqref{eq:OU}, it follows that $\lambda$ is distributed as a Gaussian variable with mean $\bar\lambda$ and autocorrelation function $\langle(\lambda(t)-\bar\lambda)(\lambda(t')-\bar\lambda)\rangle=\sigma^2 e^{|t-t'|/\tau}$ \cite{Gardiner}.  As we are interested in the interplay between the timescale of the dynamics and the environment, we keep the correlation time of the environment, $\tau$, as a control parameter throughout our analysis. Let us also note that, for sufficiently large values of $\sigma$, it may occur that $\lambda(t)<0$, so we restrict our analysis to the regime of small variability, $\sigma\ll\bar\lambda$. Numerical analyses are performed keeping the constraint that $\lambda(t)=0$ if a negative value is reached, but such events are extremely rare for $\sigma\ll\bar\lambda$.  We have verified that other forms with bounded colored noise, as for instance dichotomous Markov noise \cite{Bena2006}, do not change qualitatively our main results.

In the well-mixed scenario, substituting $\lambda\rightarrow\bar\lambda+\sigma\epsilon(t)$ in Eq. \eqref{eq:deterministic}, one readily finds the following stochastic differential equation for the averaged density $\rho$ in the infinite-size limit: \begin{eqnarray}
  \dot{\rho}(t) 
= (\bar\lambda -\mu)\rho - \bar\lambda\rho^2 + \rho(1-\rho)\sigma\epsilon(t).
\label{eq:langevin-env}
\end{eqnarray}
The set of stochastic equations formed by Eq. \eqref{eq:OU} and \eqref{eq:langevin-env} constitutes the starting point of our analysis. In this work, we exclusively focus on the impact of environmental fluctuations, as here we consider the limit of very large populations, where demographic fluctuations can be safely neglected. The important case with the combined effects of demographic and environmental stochasticity --relevant for finite systems-- has also been explored in the literature \cite{Kamenev, McKane2015, Vergassola2015, Nadav2016-1, Nadav2016-2}.

\section{Results}
We analyze how environmental colored noise changes the phase diagram and mean-extinction times (as sketched in Fig. \ref{fig:cp}) by employing both analytical and computational approaches. To this end, we combine two different analytical tools: the Unified Colored Noise Approximation (UCNA) \cite{UCNA} (see also Appendix \ref{sec:ucna}) and a path-integral approach to calculate extinction times in finite populations \cite{Kamenev}. Numerical simulations of the stochastic particle (``individual-based'') model have been implemented by means of Anderson's next reaction method \cite{Anderson}, that can be adapted to the case in which rates vary stochastically in time (see Appendix \ref{sec:anderson}).

\begin{figure}[t!] \includegraphics[width=0.95\columnwidth]{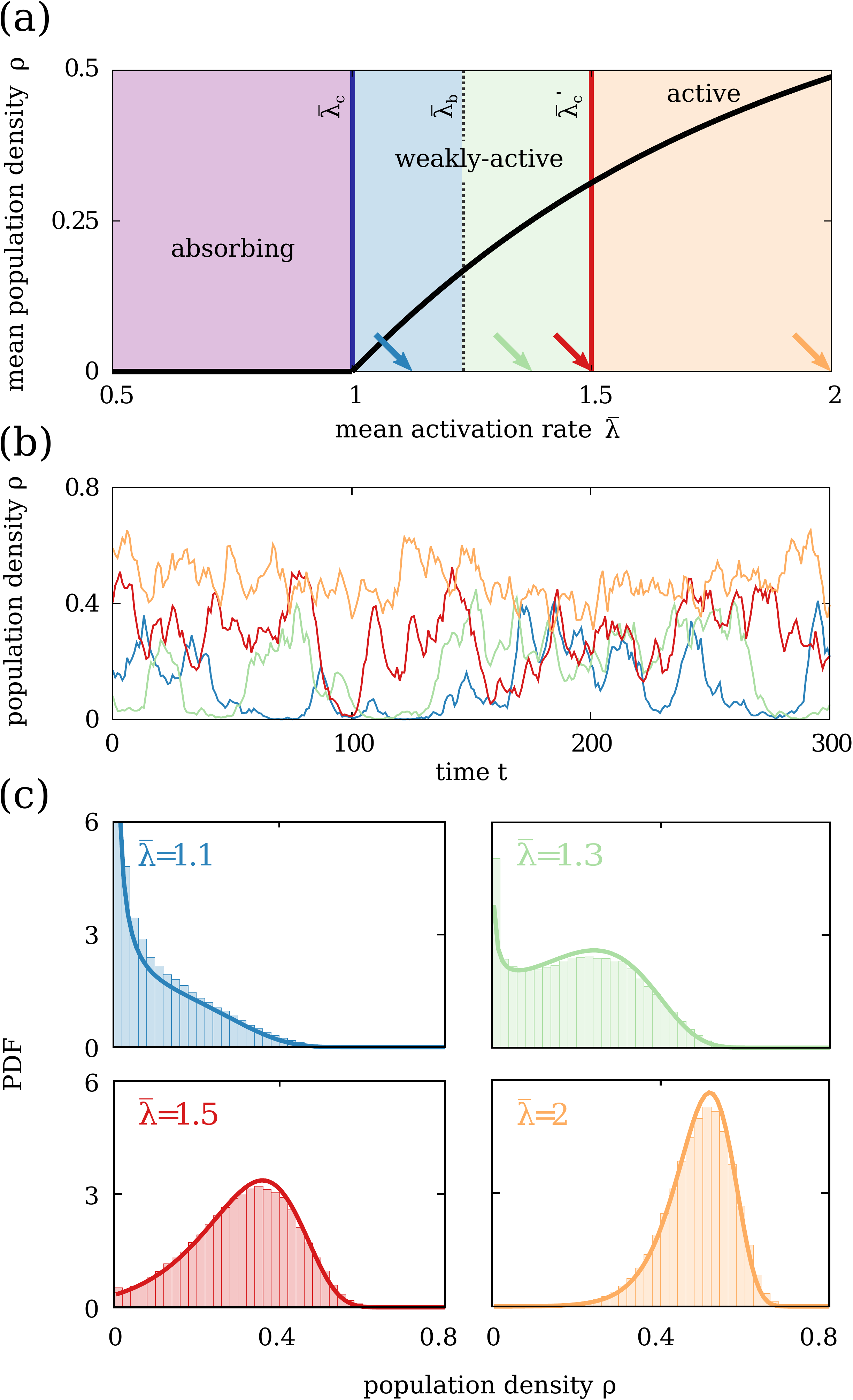} \caption{
\footnotesize
\textbf{Phases of the Contact Process with environmental noise (well-mixed scenario).} Individual based dynamics are implemented using Anderson's next reaction method \cite{Anderson}, where the death rate is fixed to $\mu=1$ and the birth rate is a stochastic Gaussian variable with mean $\bar\lambda$, variance $\sigma^2$ and temporal correlation $\tau$ (modeled as an Ornstein-Uhlenbeck process, Eq. \eqref{eq:OU}). We study the behavior of the model for different values of $\bar\lambda$, and we take $\mu=1$, $\tau=5$, $\sigma^2=0.1$, $N=1000$. Extinction is avoided by introducing an active particle at a random location in order to measure quasi-stationary distributions. \textbf{(a)} Phases of the model: i) in the absorbing phase, $\bar\lambda<\bar\lambda_c=\mu$, the only stationary solution is extinction; ii) in the weakly-active phase, $\bar\lambda_c<\bar\lambda<\bar\lambda_c' = \mu+\tau\sigma^2$, there is a positive quasi-stationary density but the system can approach arbitrarily close to extinction due to fluctuations of the environment; this phase can be divided into two phases, depending on whether the corresponding probability distribution function (PDF) is uni- or bi-modal; iii) in the active phase, excursions close to the absorbing state become extremely rare. \textbf{(c)} Timeseries of the total population density, for different values of $\bar\lambda$ (same color code than in panel a). \textbf{(c)} Histograms represent the stationary PDF of the population density in the individual-based model, whereas continuous lines are the theoretical prediction given by the UCNA approximation, Eq. \eqref{eq:UCNA-pst}. The overlap between the observed PDF $\bar{P}_{st}$ and the theoretical prediction $P_\mathrm{st}$ (computed as $\int_\rho d\rho \min\{\bar{P}_\mathrm{st}(\rho),P_\mathrm{st}(\rho)\}$) takes the values $0.90, 0.95, 0.98$ and $0.97$, for the four cases above, respectively.} \label{fig:phases} \end{figure}

\subsection{Phase diagram}
We first compute the stationary density as a function of parameter values. The process defined by Eqs. \eqref{eq:OU} and \eqref{eq:langevin-env} is Markovian and, thus, the theory of Markovian processes applies \cite{Gardiner}. The standard approach to solve it consists in finding the steady-state distribution $P_{st}(\rho,\epsilon)$ by solving the corresponding Fokker-Planck equation, and then computing its associated averaged density $\rho^*=\int_0^1 d\rho \rho \int_{-\infty}^\infty d\epsilon P_{st}(\rho,\epsilon)$.  However, this program cannot be completed analytically in the present case, as an exact integral does not exist.

The unified colored-noise approximation (UCNA) allows us to construct an \emph{approximate} Markovian process for just one variable --much more susceptible of analytical understanding-- describing the evolution of the population density with white noise \cite{UCNA}. In a nutshell, the UCNA method consists in the adiabatic elimination of the environmental variable; this can be safely done when the intrinsic dynamics and the environmental one operate at very different timescales. As a matter of fact, the method provides an exact equation for $\tau\rightarrow0$ and $\tau\rightarrow\infty$, whereas it is only approximate for intermediate timescales $\tau$. Thus, the UCNA can be understood as an ``interpolation'' between the dynamics for rapidly and slowly varying environments, respectively \cite{UCNA-interpolation} (see Appendix \ref{sec:ucna}).

For simplicity, it is convenient to rewrite Eq. \eqref{eq:langevin-env} in terms of a new variable with additive rather than multiplicative noise. In particular, defining $x=\log\left({\rho}/{(1-\rho)}\right)$, so that $x(\rho=0)=-\infty$ and $x(\rho=1)=\infty$, Eq.  \eqref{eq:langevin-env} becomes \footnote{As the noise has a finite correlation, the standard rules of calculus apply here}: \begin{equation} \dot{x}(t) = \bar \lambda - \mu -\mu e^x + \sigma \epsilon(t) \equiv f(x) + \sigma\epsilon(t), \label{eq:langevin-env2} \end{equation} where, to ease the notation we have introduced the drift term $f(x)=\bar\lambda - \mu - \mu e^x$. Unless explicitly stated, the forthcoming expressions remain valid for other choices of $f(x)$. In the case of Eq. \eqref{eq:langevin-env2}, after the elimination of the environmental variable $\epsilon$, the UCNA approximation leads to a Langevin equation with multiplicative noise of the form (see Appendix \ref{sec:ucna} for a more detailed presentation of the UCNA method): \begin{equation}
 \dot{x}(t) = a_\tau(x) + g_\tau(x)\eta(t),
 \label{eq:langevin-env2-ucna}
\end{equation}
where $\eta(t)$ is a Gaussian white noise with zero mean and $\langle \eta(t) \eta(t') \rangle=\delta(t-t')$, which has to be understood in the Stratonovich sense \cite{UCNA}, and where
\begin{equation}
 a_\tau(x) = \frac{f(x)}{1-\tau f'(x)}, \qquad g_\tau(x) = \frac{\sqrt{2\tau}\sigma}{1-\tau f'(x)}.
 \label{eq:def_FP}
\end{equation}
Eq. \eqref{eq:langevin-env2-ucna} 
is equivalent to the following  Fokker-Planck equation for the probability distribution $P(x, t)$ \cite{Gardiner}:
\begin{eqnarray}
\label{eq:UCNA-fokker-planck}
 \partial_t P(x,t) &=& -\partial_x \left[\left(a_\tau(x) +\frac{1}{2}g_\tau(x) g_\tau'(x)\right) P(x,t) \right]+\nonumber\\
 &&\frac{1}{2}\partial_x^2\left[g_\tau^2(x) P(x,t)\right]
\end{eqnarray}
whose stationary solution, $P_{st}(x)$, can be found analytically imposing the zero-flux condition:
\begin{equation}
 P_{st}(x) = Z^{-1}|1-\tau f'(x)| \exp\left[\frac{1}{\tau\sigma^2}\int^x dy  f(y) (1-\tau f'(y)) \right],
 \label{eq:UCNA-pst}
\end{equation}
where we have introduced the potential $V(x)=-\int^x f(y) dy$ and $Z$ is a normalization constant.
Introducing the expression of $f(x)$, i.e. Eq. \eqref{eq:langevin-env2}, and reverting the change of variables, $P_{st}(\rho)=P_{st}(x(\rho))dx/d\rho$,
one finally obtains the stationary distribution for the dynamics of Eq. \eqref{eq:langevin-env} under the UCNA approximation:
\begin{eqnarray}
 P_{st}(\rho) &=& Z^{-1}\rho^{\frac{\bar\lambda-\mu}{\sigma^2\tau}-1}\frac{1+(\mu\tau-1)\rho}{(1-\rho)^{\frac{\bar\lambda-\mu}{\sigma^2\tau}+2}} \times \\
 &&\exp\left[
-\frac{\mu}{\tau\sigma^2}\frac{\rho}{1-\rho} - \frac{1}{2\sigma^2}\left(\bar\lambda-\frac{\mu}{1-\rho}\right)^2
 \right].\nonumber
\label{eq:pst-rho}
\end{eqnarray}
Although Eq. \eqref{eq:pst-rho} has a complicated form, its shape is chiefly controlled by the factor $\rho^{\frac{\bar\lambda-\mu}{\sigma^2\tau}-1}$, as illustrated in Fig. \ref{fig:phases}. In particular, if $\bar\lambda<\lambda_c=\mu$, a non-integrable singularity appears at $\rho=0$, i.e. the distribution is not normalizable. This means that Eq. \eqref{eq:pst-rho} is not a truly stationary solution, and the only stationary solution corresponds to the absorbing state, $P_{st}(\rho)=\delta(\rho)$ \cite{munoz1998nature}.  On the other hand, for $\bar\lambda>\lambda_c=\mu$, Eq. \eqref{eq:pst-rho} can be safely normalized. Consequently, our first important result is that environmental variability does not shift the critical point for the process described by Eq. \eqref{eq:langevin-env}.

Remarkably, there is an important difference with respect to the case with constant rates, as the active phase splits into two regions (see Fig. \ref{fig:phases}): the first one, spanning in $\lambda_c<\bar\lambda<\lambda_c+\sigma^2\tau\equiv\lambda_c'$, for which $P_{st}(\rho=0)$ exhibits an integrable singularity, and the second one, for $\bar\lambda>\lambda_c'$, for which $P(\rho=0)=0$. Therefore, we find a region $\bar\lambda\in[\lambda_c,\lambda_c']$ where the system can be found arbitrary close to the absorbing state. We call this region the ``weakly-active'' phase. Moreover, such a region can be itself divided into two subregions; after a certain value $\bar\lambda=\lambda_{b}$ the distribution becomes bimodal, only to recover its mono-modality when $\bar\lambda>\lambda_c'$. The value of $\lambda_b$ does not have a simple analytical form and needs to be numerically determined.

We have compared the prediction of the PDF, Eq. \eqref{eq:pst-rho}, with simulations of the individual-based model.  In order to capture the form of the quasi-stationary distribution (i.e. the distribution conditioned to the fact that the system is not in the absorbing state) it suffices to instantaneously introduce one active particle in the system if the population becomes extinct (more sophisticated methods provide slightly more accurate results of the quasi-stationary distribution \cite{Oliveira2005}). Fig. \ref{fig:phases}b illustrates the timeseries of the population density for a system size $N=1000$, for different values of $\bar\lambda$ taken in the weakly-active and active phases, respectively. Histograms are represented in Fig. \ref{fig:phases}c, together with the corresponding theoretical prediction, Eq. \eqref{eq:pst-rho} (solid curves), illustrating a rather good agreement between them.

These results (summarized in Fig. \ref{fig:phases}) have been obtained neglecting the effects of demographic noise, which may play a fundamental role when the system approaches to the absorbing state. In fact, one may have doubts about the physical meaning of $\rho^*$ in the weakly-active region, as the distribution exhibits a singularity at $\rho=0$ and the system may become extinct with a relatively large probability. To shed light on this issue, in the next section we analyze the mean extinction time in the weakly-active and active regions, elucidating the meaning of the different phases in the context of finite populations.

\subsection{Mean extinction times}
\label{sec:mean-ext-time}
Mean-first passage times of a stochastic process \cite{Redner} can be computed using the framework of path integrals \cite{Kamenev-book,Assaf2010,Kamenev}. Our strategy here is to apply such a method to the ``effective'' process obtained from the UCNA approximation method, Eq. \eqref{eq:langevin-env2-ucna}.

The idea behind the path integral approach is that one can express the probability of a particular realization of the process, i.e. of a path $\{(x(t),\dot x(t))\}_t \equiv (\xx,\dot \xx)$, as
\begin{equation}
 P[(\xx,\dot\xx)] \propto \exp{\left(-S[(\xx, \dot \xx)]\right)},
\end{equation}
where $S$ in the action associated with such a path, i.e. the time-integral of the Lagrangian, $S[(\xx,\dot \xx)] = \int dt \mathcal{L}(x(t),\dot x(t))$, that encodes the dynamics we aim to describe (see below).

A particular realization of the process leading to extinction can be understood as a path passing through the state of zero density (the absorbing state), where the dynamics ceases. The leading contribution is given by the most probable path starting in the neighborhood of a deterministic attractor and ending at the absorbing state, i.e. the path $(\xx^*,\dot \xx^*)$ which obeying such constraints minimizes the action. Up to leading order (in a weak-noise, i.e. low $\sigma$, expansion) the mean time to go to extinction is then inversely proportional to the probability of such a path \cite{Kamenev}: \begin{equation} \label{eq:MET1} T \sim \exp\left(S[(\xx^*,\dot \xx^*)]\right).  \end{equation}

Let us now compute the action along the most probable path, following a standard procedure \cite{Kamenev-book,Assaf2010,Kamenev}. Given a stochastic process described by a Fokker-Planck equation (e.g. Eq. \eqref{eq:UCNA-fokker-planck}), the time evolution can be described in terms of an associated Hamiltonian operator, $\partial_t P = H P$ which, as a rule of thumb, can be written by simply identifying $-\partial_x \rightarrow p$ \cite{Assaf2010,Kamenev}. In particular, for Eq. \eqref{eq:UCNA-fokker-planck}:
\begin{equation}
 H(x,p) = p \left(a_\tau(x)+ \frac{1}{2}g_\tau(x) g_\tau'(x)\right) + \frac{1}{2} p^2 g_\tau^2(x).
 \label{eq:UCNA-h}
\end{equation}
Given that the Hamiltonian does not depend explicitly on time ($\partial H/\partial t=0$), it is a constant of motion. Moreover, as the optimal path we are looking for starts from the deterministic attractor (for which $p=0$), such a constant is equal to zero \cite{Kamenev-book}.

Imposing these constraints, one finds two solutions of $H(x^*,p^*)=0$: the trivial one, $p^*=0$, corresponding to the deterministic trajectory towards the stable attractor, and another one for $p^*\neq0$, describing the most-probable fluctuation driving the system from an initial state to the absorbing state, that, for the case of Eq. \eqref{eq:def_FP}, is: 
\begin{equation}
  p(x) = -\frac{1}{\tau \sigma^2} f(x) (1-\tau f'(x)) + \partial_x \log(1-\tau f'(x)).
  \label{eq:p-UCNA}
\end{equation}
Given that the Lagrangian is  the Legendre transform of the Hamiltonian, 
$\mathcal{L}(x,\dot x(x,p))=\dot x(x,p) p - H(x,p)$ and that $H=0$, the action can be easily evaluated as
\begin{equation}
S= \int_{t_i}^{t_f} dt \dot x(x,p) p = \int_{x_i}^{x_f} dx p(x)
\end{equation}
without the need to explicitly integrate the equations of motion (Hamilton equations).
Plugging Eq. \eqref{eq:p-UCNA} in this, and using Eq. \eqref{eq:MET1} \footnote{In the UCNA approximation we have rescaled the time variable, but this change only introduces a $\sqrt{\tau}$ factor in the mean extinction times.}, one readily obtains: \begin{eqnarray} \label{eq:MET3}
 T &\sim&
\left|\frac{1-\tau f'(x_f)}{1-\tau f'(x_i)} \right| \exp\left(\frac{1}{2\sigma^2} \left[f^2(x_f) - f^2(x_i)\right]\right) \times \nonumber\\
&&\exp\left(\frac{1}{\tau \sigma^2} \left[V(x_f)-V(x_i)\right]\right)
\label{time} ,
\end{eqnarray}
where we have defined the potential $V(x)=-\int^x dx' f(x')$. This expression, which is valid for a general form of $f(x)$, can be understood as a generalization of the \emph{Arrhenius formula} with an effective diffusion term equal to $\tau \sigma^2$.
\begin{figure}[bth]
 \includegraphics[width=\columnwidth]{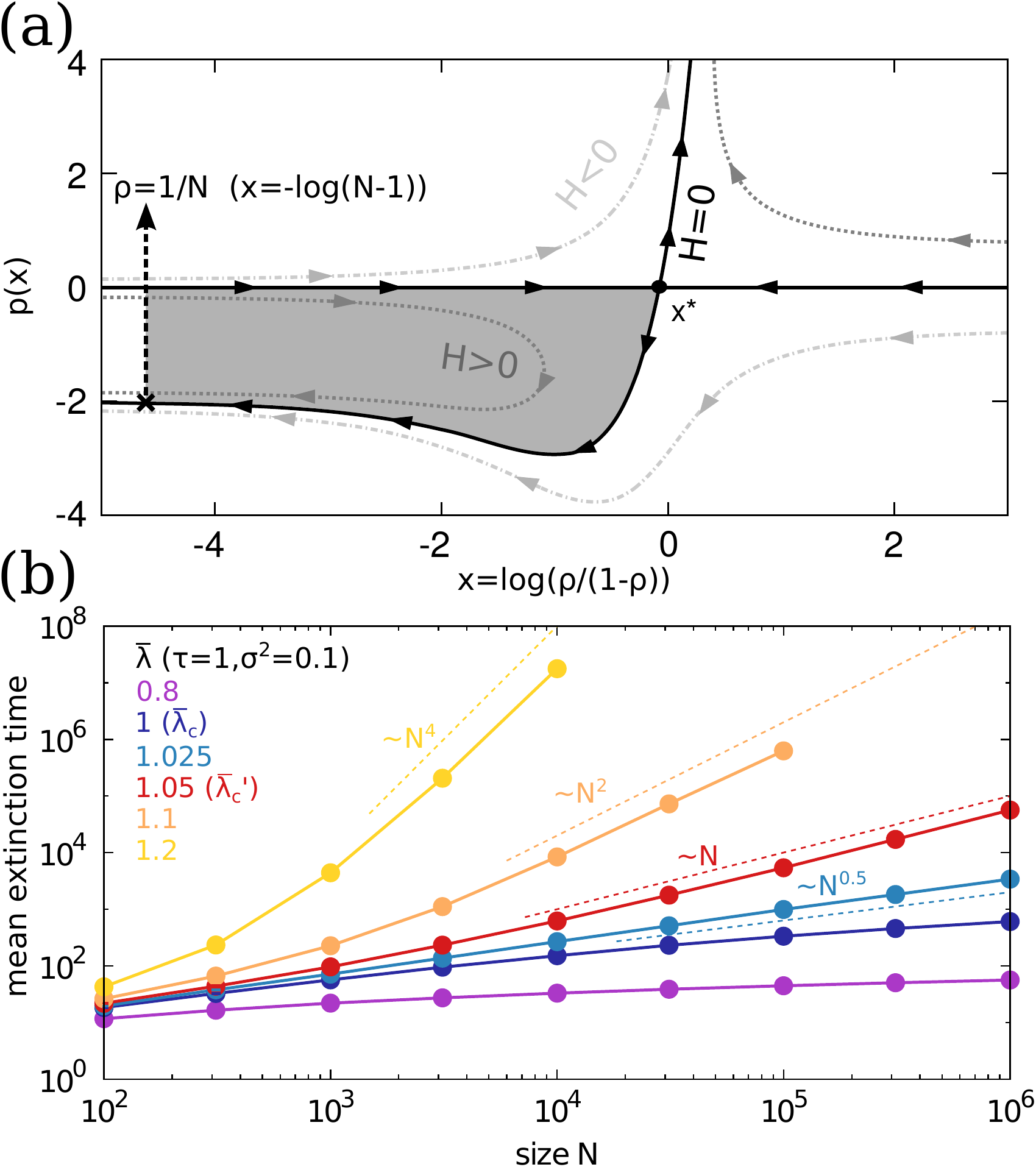}
 \caption{\textbf{(a)}: in a path-integral framework, a stochastic process follows trajectories in the $(x,p)$-space. The deterministic dynamics ($p=0$) lead the system to a stable attractor $x^*$, from where the system can escape due to fluctuations ($p\neq 0$) and go to extinction, that here we identify with the state of one single particle remaining in the population, $\rho=N^{-1} \Rightarrow x=-\log(N-1)$. Up to first order, the mean extinction time, $T$, is the exponential of the action associated to the most probable path, represented by the shaded region. \textbf{(b)} Numerical results of the individual-based model for $T$ as a function of the system size are represented with dots, together with the asymptotic theoretical behavior, Eq. \ref{eq:MET} (dashed lines), for different values of $\bar\lambda$. Barr errors are smaller than dot size. Parameters are set to $\mu=1$, $\tau=1$, $\sigma^2=0.05$. For sufficiently large values of $N$ (for which demographic fluctuations can be neglected), $T$ scales logarithmically in the absorbing phase, sub-linearly in the pre-active phase, and super-linearly in the active phase.}
 \label{fig:times}
\end{figure}
Fig. \ref{fig:times}a illustrates different trajectories in the $(x,p)$ plane for the case of the CP with environmental noise under the UCNA approximation.  Deterministic trajectories ($p=0$; i.e. horizontal axis) push the system towards the stable deterministic attractor $x^*$. On the other hand, a stochastic trajectory, i.e. ($p\neq 0$) starting arbitrarily close to the attractor, takes the system from there to the absorbing state that --in order to make it reachable in finite time-- we identify with the state with one particle remaining in the system, $\rho=1/N$ ($\Rightarrow x=-\log(N-1)$). The shaded area in Fig. \ref{fig:times} corresponds to the action $S$ of the most probable stochastic trajectory.

More quantitatively: it is possible to derive the asymptotic behavior of Eq. \eqref{eq:MET3} (i.e. its large-$N$ behavior) for the specific case of Eq. \eqref{eq:langevin-env2}. The initial point can be taken arbitrarily, as it does not depend on the system size, and the ending point scales as $x_f\simeq -\log(N)$ for large system sizes. With that, $f(x_f)=\bar\lambda-\mu+\mathcal{O}(N^{-1})$ and $f'(x_f)=\mathcal{O}(N^{-1})$. Finally, $V(x)=-(\bar\lambda-\mu) x + \mu e^x$, so that $V(x_f)=(\bar\lambda-\mu)\log(N)+\mathcal{O}(N^{-1})$, that introduced in Eq. \eqref{eq:MET3} leads to the final result: \begin{equation}
 T \sim N^{\frac{\bar\lambda-\mu}{\tau \sigma^2 }}
 \label{eq:MET}
\end{equation}
for any $\bar\lambda>\mu$. Therefore, our conclusion is that environmental noise induces power-law scaling of the mean-extinction time \textit{all along the active phase} --typical for systems under environmental stochasticity \cite{Kessler2007,TGP,TGP2,vojta2016,Hidalgo2017,Nadav2016-2}-- i.e. what has been called a \emph{temporal Griffiths phase} in the recent literature. Moreover, our result coincides exactly with the one derived by Kamenev \emph{et al.} for very large system sizes \cite{Kamenev} (i.e. when $\tau \ll \log(\tau \sigma^2 N)$ which --neglecting other parameter constants-- applies to our setting as we work in the asymptotic limit of large $N$).

Eq. \eqref{eq:MET} elucidates the meaning of the phase diagram depicted in Fig. \ref{fig:phases}: in the weakly-active phase ($\mu<\bar\lambda < \mu+\tau\sigma^2$), the system makes excursions very close to the absorbing state, and as a consequence $T$ scales \textit{sub-linearly} with the system size, whereas, in the active phase ($\bar\lambda>\mu+\tau\sigma^2$) $T$ scales \textit{super-linearly}, and extinction becomes more unlikely for large system sizes. The linear case, in between, signals a change in the convexity of the extinction-time vs. system-size curves.

We have checked the validity of Eq. \eqref{eq:MET} with an implementation of the individual-based model using Anderson's next step algorithm \cite{Anderson} (see Appendix \ref{sec:anderson}). To this end, we compute $T$ through independent realizations setting as initial condition $\rho=1/2$ and $\epsilon=0$, as a function of the system size $N$, for different values of $\bar\lambda$. As illustrated by Fig. \ref{fig:times}b, Eq. \eqref{eq:MET} perfectly captures the asymptotic scaling behavior of $T$, whereas it fails for small values of $N$, where demographic stochastic effects (not included in the above calculation) significantly affect the dynamics.

\section{Discussion}
\label{sec:discussion}

We have presented a mathematical and computational study of a simple model for a well-mixed population where the dynamics is subjected to environmental variability, consisting of a Contact Process with birth rates modeled as an Ornstein-Uhlenbeck process. Our goal was to explore how the standard phase diagram of the Contact Process is affected by the introduction of environmental noise and in particular by its temporal auto-correlations. We explored whether its critical point is shifted or not, and what the nature of the emerging phases is. For this, we choose to work in the large system size limit, so that demographic fluctuations are negligible with respect to environmental ones, and the focus was put on phases and phase transitions.

The approach presented here is simple and easy to extend to other models and consists in the successive use of two 
analytical techniques: an approximation to deal with colored noise that reduces the number of variables, and a way to compute extinction times from the resulting equation. In particular, in the mean field-limit (describing a well-mixed scenario), we employ the unified colored noise approximation \cite{UCNA} (UCNA) to replace the correlated (colored) noise by a delta-correlated Gaussian (white) noise, at the price of introducing, an effective force and an effective diffusion term in the Langevin equation describing the system. 

We verified computationally that the UCNA works remarkably well all across the phase diagram, generating steady-state density-distributions, hardly indistinguishable from the exact ones, as obtained from Monte Carlo simulations of the underlying microscopic model.  Numerical analyses were performed using a variation of the (exact) Gillespie integration method, adapted to deal with time-dependent (stochastic) rates, i.e. the so called Anderson's method \cite{Anderson}.  These analyses revealed that the probability distribution becomes a delta-Dirac at $\rho=0$ at a critical value of the averaged birth rate coinciding with the critical value for the pure contact process. In other words, the introduction of colored environmental noise does not shift the location of the critical point, separating the absorbing from the active phase.

To proceed further we applied a weak-noise approximation within a path-integral formulation of the effective white-noise problem \cite{Kamenev-book,Kamenev}.  Using this standard approach, a second important result is that in the active phase, the mean time required for the system to reach extinction scales as a power-law of the system size. This algebraic dependence of extinction times in the presence of uncorrelated environmental noise was first reported by Leigh \cite{Leigh} and was later on scrutinized by V\'azquez et al. \cite{TGP} who introduced the notion of ``temporal Griffiths phase'' (TGP) to refer to such a sort of active state. Temporal Griffiths phases are the counterpart of standard Griffiths phases \cite{vojta}, but where the role of spatial quenched disorder is played by temporal one \cite{TGP,TGP2,vojta2015,vojta2016}.

A more general study of extinction times in the presence of colored noise was elegantly tackled and solved by Kamenev \emph{et al} \cite{Kamenev}. These authors performed a path-integral formulation to the full problem, including both demographic and environmental stochasticity, and found diverse regimes depending on the ratio between system size and noise correlation time. In particular, the dependence on system-size of the mean extinction time changes from exponential in the absence of the environmental noise (as corresponds to the Arrhenius law) to a power law for a short-correlated noise (as is the case in our study) and to no dependence whatsoever for noise with very large correlation times. This last regime implies that when there are extremely long periods of adverse external conditions (as compared, using adequate rescaling units, with system size) the system reaches deterministically the absorbing state regardless its size. Let us notice that, this situation is not accessible to our approach, as we set the system in the asymptotic limit of large $N$ (where demographic effects can be neglected), and thus $\tau$ cannot be much larger than it. For the power-law regime, our simple method gives exactly the same dependence as in \cite{Kamenev}.

An interesting result of our analysis is that the active phase can be divided in two sub-phases. In the ``weakly active phase'' the probability distribution of $\rho$ has a non-vanishing value around $0$ meaning that the system makes excursions to very tiny values, at the edge of the collapse, but then it recovers. For this regime
we find that, in finite systems, extinction times scale sub-linearly with system size, while in the truly active phase, the scaling is super-linear. Note that the linear case in between signals a change of convexity in the extinction-time versus system-size curves in linear scale.

In summary, the combined use of the UCNA approximation to deal with colored noise and a standard weak-noise approximation for the resulting white noise equation allows us to construct a general approach to analyze particle systems with absorbing states in the presence of colored noise, in a relatively simple and systematic way. We believe that this combined approach should be very useful in applications in theoretical ecology, population dynamics and epidemic spreading among others.

\section*{Acknowledgments}
\label{sec:acknowledgments}
M.A.M. and J.H. acknowledge support from the Spanish MINECO Excellence project FIS2013-43201-P. J.H. acknowledges J. Grilli and A. Maritan for very useful comments.  \vspace{2cm}

\appendix

\section{Unified Colored Noise Approximation (UCNA)}
\label{sec:ucna}
We briefly review the method of the Unified Colored Noise Approximation (UCNA) \cite{UCNA}. We start from a general stochastic process with additive colored noise (see below for multiplicative noise):
\begin{eqnarray}
\dot x = f(x) + \sigma\epsilon(t)
 \label{eq:ucna1}
\end{eqnarray}
where $\epsilon(t)$ is described by an Ornstein-Uhlenbeck process (Eq. \eqref{eq:OU}). 
In order to confine the variable around a given bounded interval, we impose that $f'(x)<0$ for all values of $x$.
$\epsilon(t)$ can be eliminated from Eq.\eqref{eq:ucna1} by differentiating in time and introducing the expression for $\dot\epsilon$, Eq. \eqref{eq:OU}:
\begin{eqnarray}
 \ddot x &=& f'(x) \dot x + \sigma \dot \epsilon = f'(x) \dot x + \left( -\frac{1}{\tau} \left(\dot x - f(x)\right) + \sqrt{\frac{2}{\tau}}\sigma \eta(t) \right) \nonumber \\
	 &=& -\left(\frac{1}{\tau}-f'(x)\right)\dot x + \frac{1}{\tau} f(x) + \sqrt{\frac{2}{\tau}}\sigma \eta(t)
\end{eqnarray}
Multiplying both sides of this equation by $\tau$ and introducing a new time scale $\hat t=t/\sqrt{\tau}$, 
one obtains:
 \begin{eqnarray}
 \ddot x = -\gamma_{\tau}(x) \dot x + f(x) + \sqrt{2\tau}\sigma \eta(\sqrt{\tau} \hat t)
 \label{eq:ucna-intemediate}
\end{eqnarray}
where dots now refer to the time derivative in the scale $\hat t$, and where we have introduced the ``damping'' factor $\gamma_\tau(x)$ defined as:
\begin{equation}
 \gamma_\tau(x) = \frac{1}{\sqrt{\tau}}-\sqrt{\tau}f'(x).
 \label{eq:gamma}
\end{equation}
Finally, the stochastic term in Eq. \eqref{eq:ucna-intemediate} can be replaced by an equivalent one, $\eta(\sqrt{\tau} \hat t)\rightarrow \tau^{-1/4} \hat\eta(\hat t)$, where $\langle \hat \eta(\hat t) \hat \eta(\hat t')\rangle=\delta(\hat t - \hat t')$, as both of them have the same mean value and correlation function, $\langle \eta(\sqrt{\tau}\hat t)  \eta(\sqrt{\tau}\hat t')\rangle = \delta(\sqrt{\tau} (\hat t -
 \hat t')) = \frac{1}{\sqrt{\tau}} \delta(\hat t - \hat t') = \langle \tau^{-1/4} \hat \eta(\hat t) \tau^{-1/4} \hat\eta(\hat t') \rangle$.
 
Observe, from the definition of $\gamma_\tau$ (Eq.\eqref{eq:gamma}), that the system becomes over-damped both in the limit $\tau\rightarrow0$ and $\tau\rightarrow\infty$. Therefore, it is possible to perform an adiabatic approximation --valid for either very small or very large values of $\tau$-- by neglecting the transient contribution of the term $\ddot x$, and the process becomes approximately equivalent to the following Langevin equation with multiplicative noise: \begin{eqnarray}
\dot x = \gamma^{-1}_{\tau}(x) f(x) + \gamma^{-1}_{\tau}(x) \sqrt{2\sqrt{\tau}}\sigma \hat \eta(\hat t)
\label{eq:UCNA-langevin}
\end{eqnarray}
Let us note that this equation has to be understood in the Stratonovich sense, as the previous derivation has been carried using the rules of the standard calculus \cite{Gardiner}.  Finally, we remark that the UCNA method becomes exact in the limit $\tau\rightarrow0$ and $\tau\rightarrow\infty$, whereas it is an approximation for intermediate values of $\tau$.  Let us notice that, by Eq.\eqref{eq:gamma}, for $\tau\ll0$ the system is homogeneously over-damped (i.e. independently of $x$), whereas a dependency of $x$ is preserved for $\tau\gg1$. Therefore, we should expect that the UCNA provides accurate results independently of $x$ for short-correlated environments whereas, in highly-correlated environments, we may still find some discrepancies with numerical results for those values of $x$ for which $f'(x)$ is large.

\section{Anderson's next reaction method for stochastic time-dependent rates}
\label{sec:anderson}
The Gillespie algorithm is the most widespread method used to simulate the dynamics of discrete particle-like stochastic processes \cite{Gillespie}. Alternatively, when rates explicitly depend on time, one can use the algorithm developed by Anderson \cite{Anderson}, which can be easily adapted to the case in which such a dependency is stochastic.  In this appendix we briefly review Anderson's next reaction method,  underlining some technical issues arising from the fact that rates vary in time in a stochastic rather than deterministic way.

In short, Anderson's algorithm keeps track of different ``clocks'' counting the time remaining for each possible reaction to occur. In the ``system of reference'' of each reaction, its timer goes at unit speed; an absolute ``observer'' updates each of these timers according to their corresponding time-dependent rates, accelerating or slowing them, choosing which reaction will take place next and updating the species involved. 

At each time, the state of the system is determined by the number of members/particles of each species, $\mathbf{X}$. Each reaction $k$ is characterized by its propensity $a_k(\mathbf{X},t)$ and its state-change vector $\boldsymbol{\nu}_k$, so that $\mathbf{X}\rightarrow \mathbf{X}+\boldsymbol{\nu}_k$ when reaction $k$ occurs. The algorithm is implemented as follows: \cite{Anderson}:
\begin{enumerate}
 \item Initialize: set the number of each species, set $t=0$ and the internal clocks $P_k=T_k=0$ for each reaction $k$.
 
 \item Generate the internal firing times for each $k$, that are exponentially distributed random variable with unit mean: $P_k=-\log(r_k)$, where $r_k$ is a uniform random number in the interval $[0,1]$.

 \item Calculate propensity functions $a_k(\mathbf{X},t)$ for each $k$.
 
 \item For each $k$, find the $\Delta t_k$ for which $\int_{t}^{t+\Delta t_k} ds \, a_k(\mathbf{X},s) ds = P_k-T_k$ (see below).
 
 \item Find the minimum time-step $\Delta = \min_{k} \Delta t_k$. We denote $\alpha$ its corresponding reaction. 
 
 \item Update internal times $T_k=T_k+\int_{t}^{t+\Delta} a_k(\mathbf{X},s)$ (see below), total time $t=t+\Delta$, and  finally the number of species according to reaction $\alpha$, $\mathbf{X}=\mathbf{X}+\boldsymbol{\nu}_\alpha$.

 \item Generate a new internal firing time $P_\alpha=P_\alpha-\log(r_\alpha)$ with $r_\alpha$ a uniform random number in $(0,1)$, and go to step 3.
\end{enumerate}
A small complication arises during steps 4 and 6 when rates vary stochastically, as the method has an anticipating nature: when integrating each of the propensity functions one may integrate one or many stochastic functions up to a certain time $\Delta t_k$, and then seek for the minimum of such times, $\Delta$. In principle, this step does not require a complete knowledge of the whole stochastic trajectories at intermediate timesteps (see below). However, after this, one should re-integrate all the propensity functions up to $t+\Delta$ (step 6), but this can be a problem if we did not keep track of the stochastic trajectory with high enough precision.

We believe that this problem may be also solved using the theory of ``stochastic bridges'' \cite{bridge}, which in principle would allow us to generate an intermediate point of the stochastic path conditioned to a future value of such a process (information that afterwards could be safely erased). We prefer to not enter into this matter, and we simply keep track of the stochastic trajectories (in our case the Ornstein-Uhlenbeck process and its time integral) taking a moving time window of length $10\tau$ and precision $10^{-2}\tau$, where $\tau$ is the correlation of the environment. Intermediate points of such a discretization are calculated using the simple linear interpolation rule.

\subsection*{Updating formulas for the Ornstein-Uhlenbeck process and its time-integral}
\label{sec:ou-integral}
When implementing Anderson's next reaction method, we use the formulas derived in \cite{GillespieOU} to integrate exactly the OU process (Eq. \eqref{eq:OU}):
\begin{eqnarray}
\label{eq:updating1}
\epsilon(t+\Delta t) &=& \epsilon(t) \theta + \sigma_1 N_1\\
\label{eq:updating2}
\int_{t}^{t+\Delta t}\epsilon(s) ds &=& \epsilon(t) \tau (1-\theta) + \left(\sigma_2^2-\frac{\kappa_{12}^2}{\sigma_1^2} \right)^{1/2} N_2 + \frac{\kappa_{12}}{\sigma_1} N_1\nonumber\\
\end{eqnarray}
where $N_1$ and $N_2$ are two independent Gaussian random numbers with zero mean and unit variance, and the coefficients $\theta=\exp(-\Delta t/\tau)$, $\sigma_1^2= \sigma^2 (1-\theta^2)$, $\sigma_2^2 = 2\sigma^2 \tau^2 \left(\Delta t/\tau-2(1-\theta)+(1-\theta^2)/2\right)$ and $\kappa_{12} = \sigma^2 \tau (1-\theta^2)^2$. $\tau$ represents the temporal correlation of the process and $\sigma$ its standard deviation.  Notice that, when implementing Anderson's method, one could simply generate two random Gaussian numbers and look for the $\Delta t$ for which $\int_t^{t+\Delta t} \epsilon(s) ds = C$, where $C$ is some numerical value given by the algorithm. However, this leads to the already mentioned problem of anticipation, so it is preferable to integrate Equations \ref{eq:updating1} and \ref{eq:updating2} at regular time steps.


\end{document}